\documentclass[aps,prb,longbibliography,showpacs,twocolumn,superscriptaddress,amsmath,floatfix,amssymb,verbatim]{revtex4-2}
\usepackage[english]{babel}
\usepackage[utf8]{inputenc}
\usepackage{graphicx,amsmath,amssymb,color}
\usepackage{nicefrac}
\usepackage[titletoc,title]{appendix}
\usepackage{amsmath}
\usepackage{subfigure}
\usepackage{balance}
\usepackage[colorlinks,bookmarks=true,citecolor=blue,linkcolor=red,urlcolor=blue]{hyperref}
\usepackage[colorlinks,bookmarks=true,citecolor=blue,linkcolor=red,urlcolor=blue]{hyperref}
\usepackage{titlesec}

\hypersetup{ linkcolor=blue, citecolor=blue,  urlcolor=blue}
\DeclareUnicodeCharacter{25A1}{\textsquare}

\makeatletter

\ProvideTextCommand{\~}{LGR}[1]{\char126#1}

\@ifundefined{textcolor}{}
{%
	\definecolor{BLACK}{gray}{0}
	\definecolor{WHITE}{gray}{1}
	\definecolor{RED}{rgb}{1,0,0}
	\definecolor{GREEN}{rgb}{0,1,0}
	\definecolor{BLUE}{rgb}{0,0,1}
	\definecolor{CYAN}{cmyk}{1,0,0,0}
	\definecolor{MAGENTA}{cmyk}{0,1,0,0}
	\definecolor{YELLOW}{cmyk}{0,0,1,0}
}

\makeatother

\makeatother

\begin{document}

\title{Magnetic {properties} of a $J_\textnormal{eff} =1/2$ based frustrated triangular lattice antiferromagnet {Ba$_4$YbReWO$_{12}$}}

\author{M. Barik}
\affiliation{Department of Physics, Indian Institute of Technology Madras, Chennai 600036, India}
\author{J. Khatua}
\affiliation{Department of Physics, Sungkyunkwan University, Suwon 16419, Republic of Korea}
\author{Suyoung Kim}
\affiliation{Department of Physics, Simon Fraser University, Burnaby, BC V5A 1S6, Canada
}
\author{Eundeok Mun}
\affiliation{Department of Physics, Simon Fraser University, Burnaby, BC V5A 1S6, Canada
}
\author{Suheon Lee}
\affiliation{Center for Artificial Low Dimensional Electronic Systems, Institute for Basic Science, Pohang 37673, Republic of Korea}
\author{Bassam Hitti}
\affiliation{Centre for Molecular and Materials Science, TRIUMF, Vancouver, British Columbia, Canada V6T 2A3
}
\author{Gerald D. Morris
}
\affiliation{Centre for Molecular and Materials Science, TRIUMF, Vancouver, British Columbia, Canada V6T 2A3
}
\author{Kwang-Yong Choi}
\affiliation{Department of Physics, Sungkyunkwan University, Suwon 16419, Republic of Korea}
\author{P. Khuntia}
\email{pkhuntia@iitm.ac.in}
\affiliation{Department of Physics, Indian Institute of Technology Madras, Chennai 600036, India}
\affiliation{Quantum Centre of Excellence for Diamond and Emergent Materials,
Indian Institute of Technology Madras, Chennai 600036, India}

%%-------------------------------
%               ABSTRACT
%%-------------------------------

\begin{abstract}
The subtle interplay between competing degrees of freedom, crystal electric fields, and spin correlations can lead to exotic quantum states in 4\textit{f} ion-based {frustrated triangular lattice antiferromagnets}. We present the crystal structure, thermodynamic and muon spin relaxation ($\mu$SR) studies of the 4\textit{f} ion-based frustrated magnet Ba$_4$YbReWO$_{12}$, wherein Yb$^{3+}$ ions constitute a triangular lattice. The magnetic susceptibility does not show any signature of spin freezing down to 1.9 K or long-range magnetic ordering down to 0.4 K. The low-temperature Curie-Weiss fit to the inverse magnetic susceptibility data reveals a weak antiferromagnetic exchange interaction {, which is corroborated by the fit of magnetic specific heat  data following the $J_1-J_2$ model with the nearest neighbor exchange interaction of $J_1$ $\approx -0.197$ K} between the \textit{J}$_\textnormal{eff}$ = 1/2 states of the Yb$^{3+}$ moments in the lowest Kramers doublet. The lowest Kramers ground state doublet is well separated from the first excited state with a gap of $\Delta_\textnormal{CEF}$ = 278 K, as evidenced by our $\mu$SR experiments that support the realization of $J_\textnormal{eff} = 1/2$ at low temperatures. The specific heat  {experiments do not detect a phase transition down to 56 mK. The magnetic specific heat shows  a broad maximum  ~90 mK suggesting a disordered ground state with short range spin correlations.} The associated {magnetic} entropy release at low temperatures is consistent with that expected for the \textit{J}$_\textnormal{eff}$ = 1/2 state. The zero-field $\mu$SR measurements show neither the signature of spin freezing nor a phase transition, at least down to 43 mK. Our results suggest a {dynamic, disordered } ground state in this \textit{J}$_\textnormal{eff}$ = 1/2  frustrated triangular lattice antiferromagnet. Ba$_4$\textit{R}ReWO$_{12}$ (\textit{R}=rare earth) offers a viable platform to realize intriguing quantum states borne out of spin-orbit coupling and frustration.
\end{abstract}

\date{\today}
%\pacs{75.50.Bb, 75.50.Cc, 61.43.-j, 85.75.-d, 31.15.A}

\maketitle

\section{Introduction}
Frustrated magnets, {owing to} the incompatibility of competing exchange interactions between localized magnetic moments in a spin-lattice, {can} lead to a macroscopically degenerate ground state manifold and exotic quantum states such as quantum spin liquid. A quantum spin liquid is an exotic state of matter {described} by frustration-induced strong quantum fluctuations that melt {long range} magnetic order down to absolute zero temperature, deconfined fractional excitations, and long-range quantum entanglement \cite{khatua_experimental_2023,lacroix_introduction_2011,diep_frustrated_2020,khuntia_spin_2016,khuntia_gapless_2020,arh_ising_2022}. Notably, frustrated magnets are ideal hosts of a plethora of non-trivial quantum states, including non-collinear magnetic ordering, quantum criticality, magnetization plateau, Bose-Einstein condensation and topological phase transition, due to the intricate interplay between frustration, quantum fluctuation and spin correlations \cite{khatua_experimental_2023,lacroix_introduction_2011,diep_frustrated_2020, sachdev_quantum_2011,lee_interplay_2010-1}. {In addition, frustrated quantum magnets are highlighted as promising candidates for potential applications in quantum technologies. For instance, highly entangled fractional excitations in spin liquids can be exploited for fault-tolerant qubits} \cite{satzinger_realizing_2021}. {The} triangular spin-lattice stands out as {the simplest} model system for the realization of such exotic quantum states. The ground state in these {quantum magnets} is {dominated} by {an intriguing} interplay among geometrical frustration, exchange interactions beyond nearest neighbors, magnetic anisotropy, crystal electric fields, and spin-orbit coupling \cite{balents_spin_2010,kamiya_nature_2018,haley_half-magnetization_2020, li_kosterlitz-thouless_2020, yamashita_gapless_2011,yunoki_two_2006,tu_gapped_2024,PhysRevB.93.140408}. In this {context}, systems particularly those with spin \textit{S} = 1/2 are of utmost importance as the low magnetic moment induces strong quantum fluctuation {as per Heisenberg's uncertainty principle}, preventing the system from undergoing conventional long-range ordering \cite{norman_colloquium_2016}.

While the material realization of the triangular lattice quantum magnets has been extensively explored in Cu$^{2+}$-based systems\cite{shimizu_emergence_2006,yuan_possible_2022,zhou_spin_2011,zheng_two-dimensional_2018}, {an alternative platform is offered by effective spin ($J_\textnormal{eff}$) $=1/2$ as a result of the interplay between the spin-orbit coupling (SOC) and the crystal-electric field (CEF) in ions with SOC} \cite{cao_physics_2021}. In this context, the energy levels of octahedrally coordinated $\textnormal{Yb}^{3+}$ ($4f^{13}$) and $\textnormal{Ce}^{3+}$ ($4f^1$) are favorable for the formation of a distinctly-separated lowest Kramers doublet state owing to their low-symmetry crystal electric field, resulting in a significantly reduced magnetic moment at low-temperatures \cite{sichelschmidt_effective_2020, khatua_magnetic_2022}. The reduced exchange interaction attributed to the strongly localized 4\textit{f} orbitals {offers a route} to explore ground states driven by weak interactions such as dipolar interaction which may lead to elusive magnetic monopole excitations as predicted in frustrated magnets including Yb$_2$Ti$_2$O$_7$ \cite{chang2012higgs} and dipolar spin liquid as observed in Ba$_3$Yb(BO$_3$)$_3$\cite{bag_realization_2021,zeng_nmr_2020}. The weak exchange interaction between 4\textit{f} moments, leading to a substantially lower transition temperature, combined with  frustration-induced suppression of magnetic entropy, {may have} potential applications in magnetocaloric effect \cite{tokiwa_frustrated_2021}.

Geometrically frustrated magnets based on 4\textit{f} Kramers ions provide an ideal ground for realizing intriguing quantum phenomena \cite{gao_experimental_2019, tokiwa_possible_2016, feng_bose-einstein_2023, dun_yb_2013, scheie_proximate_2024}. The CEF surrounding the Kramers ions modifies the anisotropy and consequently, the ground state. For instance, the Ising anisotropy in a triangular lattice antiferromagnet NdTa$_7$O$_{19}$ stabilizes a QSL state \cite{arh_ising_2022}, {whereas easy-plane anisotropy, evident from the anisotrpic \textit{g}-factor,} in NaYbS$_2$ \cite{baenitz_naybs_2018} and  NaYbO$_2$ \cite{ranjith_field-induced_2019,grusler_role_2023}, in conjuction with frustration {induced quantum fluctuations lead to} field-induced quantum phase transitions and metamagnetic transitions such as 1/3 plateau and 1/2 plateau originating from up-up-down and up-up-up-down configurations, respectively \cite{ bordelon_spin_2020,nuttall_quantitative_2023,zangeneh_single-site_2019}. These quantum phase transtions can be tuned by external control parameters such as magnetic field \cite{feng_bose-einstein_2023} and pressure \cite{kermarrec_ground_2017}, leading to perturbation induced states. However, the presence of disorder and lattice-imperfections poses a significant challenge in the realization of such quantum phenomena, as they can substantially modify the ground state properties. For instance, in the widely studied triangular lattice antiferromagnet YbMgGaO$_4$, {strong next nearest neighbor} Heisenberg interaction is expected to stabilize a stripe-ordered state in the absence of Ga$^{3+}$/Mg$^{2+}$ antisite disorder, but eventually prevents from undergoing any long-range ordering \cite{PhysRevLett.119.157201,li_rare-earth_2015,li_crystalline_2017}. Further this disorder induced randomness in the magnetic moment distribution \cite{khatua2022signature},  structural \cite{PhysRevB.101.020406} and electrostatic disorder \cite{PhysRevLett.119.157201},  obscure the intrinsic ground state. While weak unavoidable antisite disorder is often expected to drive the system to a spin-glass phase or those with higher site dilution to random singlet state \cite{khatua2022signature}, candidate materials where the magnetic ground state {results} from the influence of disorder, are of importance for advancing the theoretical understanding of the percolation threshold for sustaining the {magnetic order parameter} in such systems \cite{vojta2006rare}. Understanding the magnetic properties of such systems may provide insights into intriguing phases borne out of the interplay between electrostatic disorder, CEF effects, anisotropy, frustration induced unconventional low-energy excitations and dipolar interaction \cite{khatua_experimental_2023,jena2025nature}.

\begin{figure*}[hbt]
    \centering
    \includegraphics[width=1\linewidth]{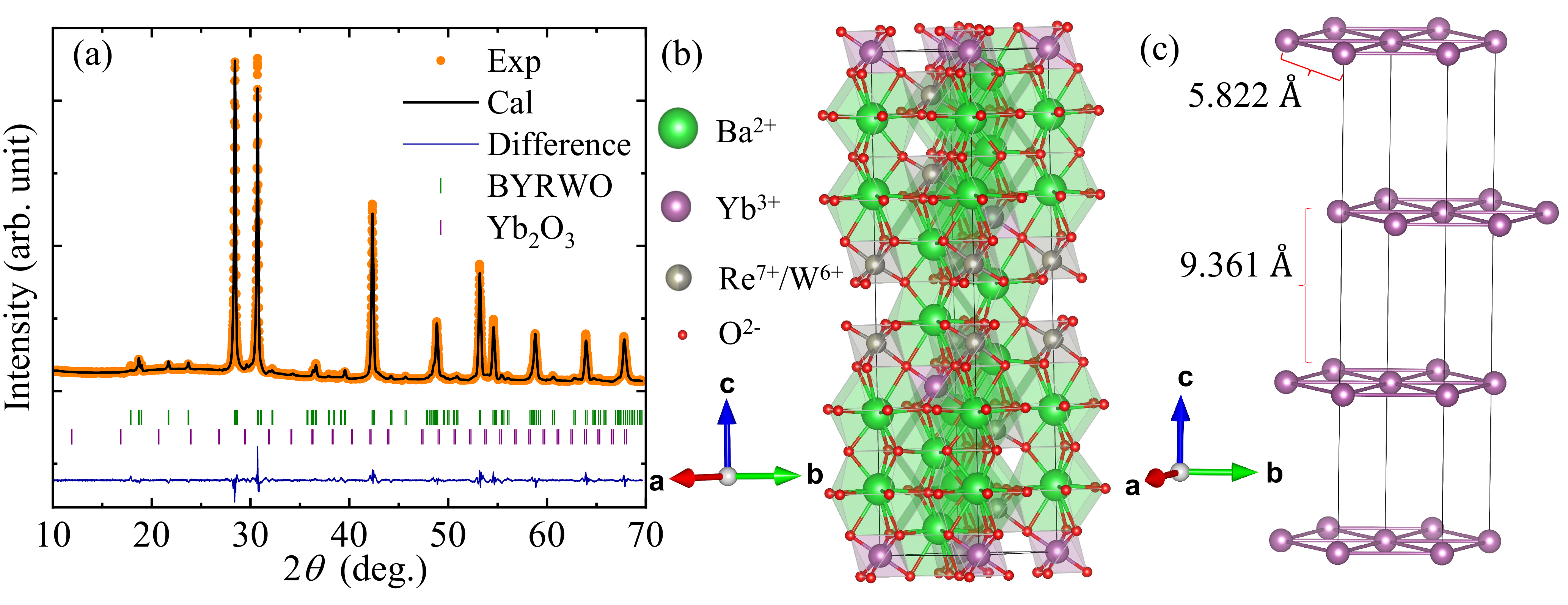}
    \caption{(a) {Two-phase} Rietveld refinement of the XRD pattern of the polycrystalline sample of Ba$_4$YbReWO$_{12}$, recorded at room temperature, indicates that it crystallizes in trigonal crystal structure with $R\Bar{3}m$ space group. The solid orange circles represent the experimental data, the black line is the calculated pattern, the green vertical bars are the Bragg's positions and the blue line indicates the difference between the experimental and simulated curve. Minor impurity observed at 2$\theta$ 
 $= $ 29.9$^\circ$ is refined by including 2.3 \%~Yb$_2$O$_3$ phase, indicated by violet vertical bars. (b) Schematic of one unit cell of BYRWO generated by VESTA. The YbO$_6$ octahedra (violet) are connected to BaO$_{12}$ polyhedra (green) and (Re/W)O$_6$ octahedra (grey). (c) Layers of Yb$^{3+}$ triangles are stacked along the c-axis.}
    \label{fig:1}
\end{figure*}

Herein, we have investigated Ba$_4$YbReWO$_{12}$ (henceforth BYRWO), which crystallizes in $R\Bar{3}m$ space group, where the Yb$^{3+}$ ions {constitute} triangular lattices. The magnetization measurements reveal the absence of long-range ordering down to 400 mK. From the Curie-Weiss fit to the inverse susceptibility, the estimated Curie-Weiss temperature that is related to the average exchange interaction, is found to be {$-0.25$ K}. The low value of effective magnetic moments {compared to that of free Yb$^{3+}$ ion value} is an indication of the realization of the lowest Kramers doublet ground state with $J_\textnormal{eff}=1/2$ state at low temperatures, that is supported by magnetic entropy expected for \textit{J}$_\text{eff}=1/2$. The $\mu$SR experiments reveal that the  lowest Kramers doublet ground state is well separated from the first  excited state with a gap of 278 K that confirms a \textit{J}$_\text{eff}=1/2$ state at low temperatures. Despite the presence of unavoidable Re$^{7+}$/W$^{6+}$ disorder, BYRWO {does not host a spin-freezing state  as evidenced by $\mu$SR down to 43 mK. The specific heat rules out the presence of long range magnetic ordering down to 56 mK. Magnetic specific heat in zero field reveals the presence of short-range spin correlations, as indicated by a broad maximum at $\approx$ 90 mK. {The zero field magnetic specific heat is reproduced with the $J_1-J_2$ model that yields a weak antiferromagnetic nearest neighbor interaction $J_1\sim -0.197$ K between $J_\text{eff} = 1/2$ moments at low temperatures.}}  Muon spin relaxation experiments  neither detect spin freezing nor long range magnetic ordering down to 43 mK. {Our thermodynamic and $\mu$SR experiments demonstrate a dynamic ground state down to 43 mK, which suggests a disordered ground state induced by frustration in this \textit{J}$_\text{eff} = 1/2$ triangular lattice antiferromagnet.}

\section{Experimental details}
Polycrystalline materials of BYRWO were synthesized via solid-state reaction method with the stoichiometric mixture of $\textnormal{BaCO}_3$ (Alfa Aesar, 99.997$\%$), $\textnormal{Yb}_2\textnormal{O}_3$ (Alfa Aesar, 99.998$\%$), Re (Alfa Aesar, 99.997$\% $) and $\textnormal{WO}_3$ (Alfa Aesar, 99.998$\%$). Prior to the synthesis, $\textnormal{BaCO}_3$ was dried overnight at $100 ^\circ$C to remove preabsorbed moisture and $\textnormal{Yb}_2\textnormal{O}_3$ was preheated at $800 ^\circ$C to remove carbonates. The reactants mixture was pelletized and annealed at $1100^\circ$C for 24 hrs to achieve the desired product \cite{kemmlersack_uber_1979}. To confirm the phase of the formed sample, we conducted X-ray diffraction measurements of the polycrystalline sample in Aeris PANalytical X-ray diffractometer using Cu-$\textnormal{K}_\alpha 
 $ $(\lambda=1.54$ \AA) radiation. Magnetization measurements were carried out by employing a vibrating sample magnetometer (VSM) option using a Quantum Design(QD), physical properties measurement system (PPMS) in the range 1.9 K$\leq T \leq$ 300 K. The measurements down to 0.4 K were performed by using a Helium-3 option to superconducting quantum interference device (SQUID) magnetometer in the field range 0 T $\leq \mu_0H \leq$ 7 T. To record the specific heat in the temperature range 1.9 K $\leq T \leq $ 200 K with the application of the field 0 T $\leq \mu_0H \leq$ 7 T , the thermal relaxation method, implemented in a PPMS, was used. For the measurement of the specific heat in the range 0.4 K $\leq T \leq$ 4 K and 0 T $\leq \mu_0H \leq$ 9 T a Helium-3 option and for 0.056 K $\leq T \leq$ 3.9 K under the application of 0 T $\leq \mu_0H \leq$ 3 T, a dilution refrigerator, equipped with the PPMS was employed.

Muon spin relaxation ($\mu$SR) is an excellent probe to shed microscopic insights into the ground state in frustrated magnets. Zero-field (ZF) and longitudinal field (LF) muon spin relaxation ($\mu$SR) measurements were conducted using the M15 and M20 beamlines at TRIUMF in Vancouver, Canada. The M15 station was equipped with a dilution refrigerator, and Ba$_{4}$YbReWO$_{12}$ powder samples compressed into pellet were mounted on a silver cold finger. To ensure efficient heat transfer between the sample and the cold finger, a mixture of copper grease and Apiezon N grease was used, and the sample was wrapped in a thin layer of silver foil. ZF $\mu$SR measurements were performed in the  temperature range 0.036 K $\leq$ $T$ $\leq$ 4 K, and additional measurements were carried out at 36 mK under different longitudinal fields at the M15 beamline. At the M20 beamline, approximately 0.5 g of the powder samples were placed into a thin envelope made of Mylar tape coated with aluminum, approximately 50 µm thick. This envelope was then mounted on a copper fork sample stick to facilitate the measurements. A standard $^4$He flow cryostat was used to achieve a base temperature of 2 K at the M20 beamline, ensuring precise control over the experimental conditions throughout the temperature range  2 K $\leq$ $T$ $\leq$ 250 K. In these $\mu$SR experiments, 100\% spin-polarized muons were implanted into the sample, and the time evolution of the muon spin polarization was tracked by monitoring the asymmetric distribution of positrons, detected by forward and backward detectors positioned on either side of the sample \cite{hillier_muon_2022}. {For each temperature, we collected $\mu$SR spectra with a total of approximately 20 million counts to ensure high statistical accuracy.} The collected $\mu$SR spectra were analyzed using the MUSRFIT software  package \cite{suter_musrfit_2012}.

\vspace{-\baselineskip}
\section{Results}
\subsection{Structural details}
To ascertain the phase composition of the final product and to extract crystallographic parameters, we conducted the Rietveld refinement \cite{rietveld_profile_1969} of the XRD pattern acquired at room temperature by using FullProf suite \cite{rodriguez-carvajal_recent_1993} with reference to the isostructural compound $\textnormal{Ba}_4\textnormal{ScReWO}_{12}$ \cite{kemmlersack_uber_1979}. Alongside the predominance of  BYRWO phase, a small fraction of unreacted $\textnormal{Yb}_2\textnormal{O}_3$, which is unavoidable in some polycrystalline samples and less likely to have substantial influence on the intrinsic magnetic properties of the material \cite{khatua_magnetic_2024, grusler_role_2023,tokiwa_frustrated_2021}, was calculated to be around {2.4} \% from the Rietveld refinement.
\begin{table*}[hbt]
     \centering
     \caption{Structural parameters obtained from Rietveld refinement of XRD data acquired at room temperature. BYRWO crystallizes in  $R\Bar{3}m$ space group with the unit cell parameters  $a=b= 5.8227(3)$ \AA, $c= 28.083(3) $ \AA, $\alpha = \beta = 90^\circ$, and $\gamma= 120^\circ$. The goodness of fit parameters for the refinements are obtained as {$R_\textnormal{p} =13.3 $, $R_\textnormal{wp} =11.2$, $R_\textnormal{exp} = 4.24$ and $\chi^2 =7.02$.}}
     
     \begin{tabular}{cccccccc}
     \hline
     \hline
  
     ~~~~~~~~~~ Atom ~~~~~~~ & ~~~~~~~~~~label~~~~~~~~~~~ &Wyckoff positions~~~~~~~& ~~x & ~y &~~z &~~~~Occ.&~~~~~~~~ {B$_{iso}$}~~~~~~~ \\
      \hline
       Yb   & Yb & 3a & 0.000 & 0.000 & 0.000 & 1 & {0.443}\\
       Ba   & Ba1 & 6c & 0.000 & 0.000 & 0.2745(3) & 1 & {0.553}\\
       Ba   & Ba2 & 6c & 0.000 & 0.000 & 0.1414(3) & 1 & {0.232}\\
       W   & W & 6c & 0.000 & 0.000 & 0.434(1) & 0.5& {0.096}\\
       Re   & Re & 6c & 0.000 & 0.000 & 0.434(1) & 0.5& {0.129}\\
       O   & O1 & 18h & 0.14(3) & 0.35(8) & 0.6055(1) & 1& {0.640}\\
       O   & O2 & 18h & 0.183 & 0.3666(1) & 0.463(5) & 1& {0.99}\\
            \hline
     \hline
     \end{tabular}

     \label{tab:1}
   
 \end{table*}
\begin{figure*}[hbt]
    \centering
    \includegraphics[width=1\linewidth]{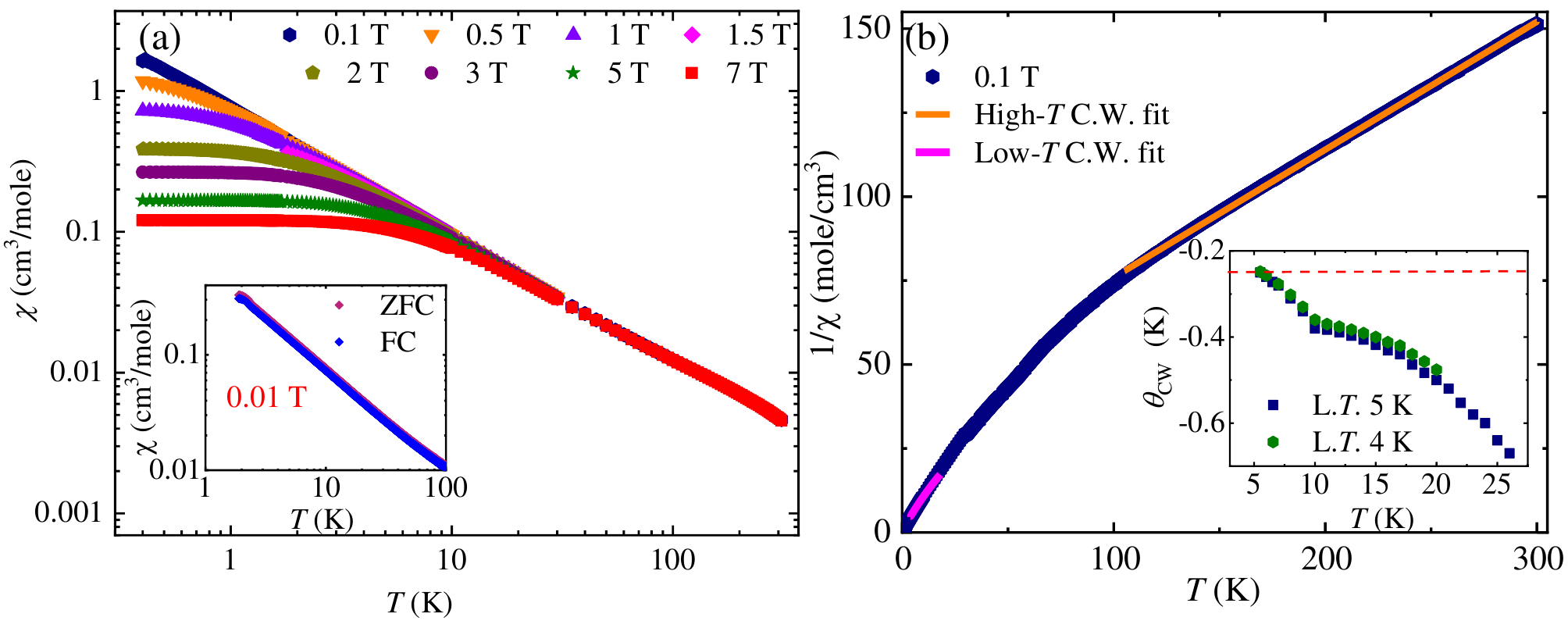}
    \caption{(a) Temperature dependence of magnetic susceptibility of BYRWO obtained in various magnetic fields down to 0.4 K. The inset shows the absence of bifurcation of zero field-cooled (ZFC) and field-cooled (FC) susceptibilities recorded at 100 Oe. (b) The Curie-Weiss fits to the inverse susceptibilities. The  solid orange line represents the high-temperature CW fit and the solid pink line is the low-temperature CW fit. {The inset shows the variation of $\theta_\textnormal{CW}$ by changing the upper limit of the fit with the lower limit being fixed at 4 K and 5 K}.}
    \label{fig:2}
\end{figure*}

Figure \ref{fig:1}(a) depicts the two-phase Rietveld refinement of the XRD pattern, revealing that BYRWO crystallizes in a trigonal structure with space group $R\Bar{3}m$ with lattice parameters $a=b= 5.8227(3)$ \AA, $c= 28.083(3) $ \AA, $\alpha = \beta = 90^\circ$, and $\gamma= 120^\circ$. The refined atomic coordinates and their respective occupancies are presented in Table \ref{tab:1}. The crystalline arrangement generated by Visualization for Electronic and Structural Analysis (VESTA) software is illustrated in Fig. \ref{fig:1}(b) \cite{momma_vesta_2011}. The adjacent intraplanar Yb$^{3+}$ ions are decorated on triangular spin lattices with Yb-Yb bond-length $\approx 5.822$ \AA, wherein the superexchange interaction between the YbO$_6$ octahedra is mediated via BaO$_{12}$ polyhedra following Yb-O-Ba-O-Yb pathway and/or (Re/W)O$_6$ octahedra with the pathway Yb-O-(W/Re)-O-Yb. The four layers of Yb-triangles, stacked at equal spacing of 9.35~\AA~from each other along the \textit{c}- axis ensure the vanishing small interplanar exchange interaction which is likely to be mediated through elongated Yb-O-Ba-O-Ba-O-Yb pathway as the interaction pathway via $\textnormal{Re}^{7+}/\textnormal{W}^{6+}$  is longer. However, the unavoidable $\textnormal{Re}^{7+}/\textnormal{W}^{6+}$ site disorder due to similar ionic radii, modifies the charge environment surrounding $\textnormal{Yb}^{3+}$, which may lead to bond-disorder \cite{kimchi_valence_2018}. The structure of BYRWO is similar to YbMgGaO$_4$, which also crystallizes with the $R\Bar{3}m$ space group, with $\textnormal{Mg}^{2+}/\textnormal{Ga}^{3+}$ charge disorder in between the $\textnormal{Yb}^{3+}$ triangular planes \cite{li_rare-earth_2015}. However, unlike the direct Yb-O-Yb linkage in YbMgGaO$_4$, BYRWO exhibits a more indirect in-plane exchange pathway and significantly longer Yb–Yb bond lengths (5.82~\AA ), which together account for the reduced exchange interaction in BYRWO.

\raggedbottom
The $c/a$ ratio in BYRWO, a rough estimate of the ratio of the exchange interaction between the out-of-plane and in-plane, is $1.61$, less than $\approx2.47$ in YbMgGaO$_4$, but more than in triangular lattice materials NaBaYb(BO$_3$)$_2$ $\approx 1.09$ \cite{guo_magnetism_2019} and Ba$_6$YbTi$_4$O$_{17}$ $\approx 1.26$ \cite{khatua_magnetic_2024}. The comparison of the interplanar to intraplanar distances of other Yb-based triangles along with their magnetic ground states are shown in Table \ref{tab:2}.

\begin{figure*}[hbt]
    \centering
    \includegraphics[width=1\linewidth]{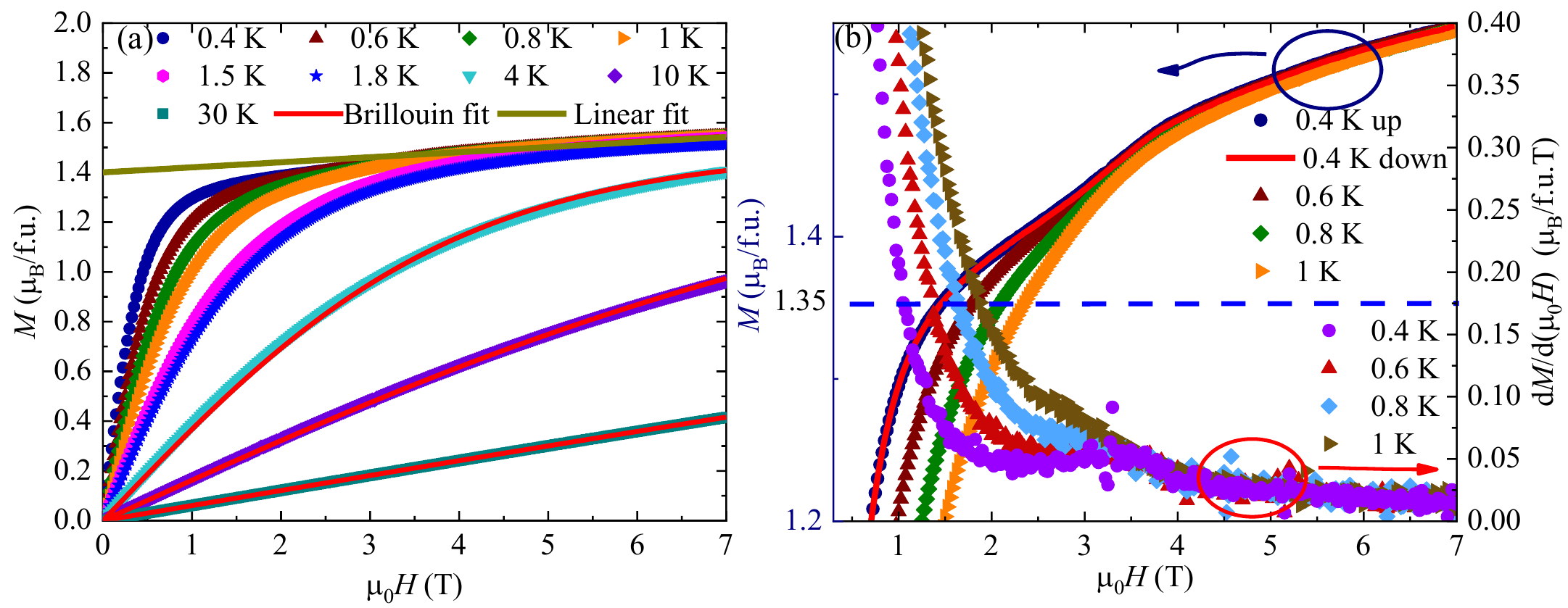}
    \caption{(a) Magnetization isotherms obtained at various temperatures. The solid red lines depict the fits of the Brillouin function. {The solid dark yellow line represents the linear fit to the high field magnetization in order to extract Van Vleck contribution from the slope.} (b) Left: Zoomed panel of the magnetic isotherms below 1 K, where small plateau like behavior is observed, Right:Variation of $dM/d(\mu_0H)$ with the field.}
    \label{fig:3}
\end{figure*}

\vspace{-\baselineskip}
\subsection{Magnetization}
The magnetic susceptibilities measured in the range $0.4$ K $\leq$ $T$ $\leq$ $300$ K under the application of magnetic fields (0.1 T $\leq\mu_0B\leq $ 7 T) are depicted in Fig. \ref{fig:2}(a). The absence of any anomaly in the magnetic susceptibility down to 0.4 K indicates that there is no long-range ordering. The inset shows the overlapping of zero field-cooled (ZFC) and field-cooled (FC) susceptibilities obtained at 0.01 T down to 1.9 K, suggesting the absence of frozen moments or ferromagnetic nature of the interactions. In order to extract the Curie-Weiss temperature ($\theta_\textnormal{CW}$), which is related to the exchange interaction strength, and the effective magnetic moment ($\mu_\textnormal{eff}$), the inverse susceptibility at 0.1 T was fitted with the Curie-Weiss law $\chi = {\chi_0}+C/(T-\theta_\textnormal{CW})$, where $C$ is the Curie constant related to the effective moment by $\mu_\textnormal{eff} \approx \sqrt{8C}$ and $\chi_0$ {is the temperature independent susceptibility, which is the summation of the core diamagnetic susceptibility of each ion and Van Vleck paramagnetic susceptibility.} Unlike 3$d$, 4$d$ and 5$d$ based systems, for 4$f$ systems, $\theta_\textnormal{CW}= -99.5 \pm 0.1$ K from the Curie-Weiss fit in the high-temperature region indicates the thermal population of CEF excitations, rather than the exchange strength \cite{arh_ising_2022,khatua_spin_2022}. The obtained $\mu_\textnormal{eff}$ $=$ 4.57 $\mu_\textnormal{B}$ is close to the expected magnetic moment of the free Yb$^{3+}$ ions. The inverse susceptibility below 100 K deviates from the linearity due to the crossing of the energy levels of the CEF of Yb$^{3+}$ corresponding to the temperature. The Yb$^{3+}$ ion ($^2F_{7/2}$, $L=3$, $S=1/2$ and $J=7/2$) surrounded by the octahedral crystal electric field leads to eight energy levels and four Kramers doublet states with $J_\textnormal{eff} =1/2$ ground state, which is evident from the reduced effective moment $\mu_\textnormal{eff} ~   {\approx 2.62} $
 $\mu_\textnormal{B}$ obtained from the low-temperature CW fit, consistent with other Yb-based systems \cite{khatua_magnetic_2022,sana_magnetic_2023}. The Land\'e \textit{g} factor (\textit{g}) obtained from the $\mu_\textnormal{eff}$, by fixing $J_\textnormal{eff} = 1/2$ turns out to be {$\sim$ 3.02,} which is comparatively larger than that of a free spin, owing to the presence of spin-orbit coupling. {In order to estimate the low temperature $\theta_{CW}$ precisely, the upper limit of the temperature range was varied till 26 K with the increment of 1 K, while the lower limit was fixed to be 5 K and 4 K \cite{arh_ising_2022}. By varying the low-temperature from 5 K to 4 K, there is no significant change in $\theta_{CW}$, indicating the reliability of the C.W. temperature \cite{arh_ising_2022}.} {In addition, the temperature independent susceptibility, was subtracted from the total  susceptibility. {The core diamagnetic susceptibility was calculated from the summation of each constituent ion.} The Van Vleck susceptibility was obtained from the linear fit to the high field part of the magnetization isotherm at 0.4 K (see Fig. \ref{fig:3}(a)). The slightly larger value of Van Vleck susceptibility of 0.0113 cm$^3/$mole results from the presence of significant number of orphan spins due to the Re$^{7+}$-W$^{6+}$ disorder and the CEF levels of Yb$^{3+}$ ion. The negative value of $\theta_\textnormal{CW}\sim$ $-0.25$ $\pm$ 0.01 K} {obtained after subtracting $\chi_0$} indicates the presence of a weak antiferromagnetic exchange interaction between the Yb$^{3+}(J_\text{eff}=1/2)$ moments at low temperatures. From the mean-field approximation, the exchange strength $J$ between the nearest neighbor spins obtained from the CW temperature, is related by $J/k_\textnormal{B}=3\theta_\textnormal{CW}/{zS(S+1)}$, where \textit{z} denotes the coordination number of each Yb$^{3+}$ moments, which is 6 for triangular lattices. Using this relation, the average exchange strength between $J_\text{eff}=1/2$  moments is estimated to be $J/k_B \approx {-0.166}$ K, which is similar to that found in other Yb-based triangular lattices with longer exchange path \cite{khatua_magnetic_2022,khatua_magnetic_2024}. {This is further supported by reproducing the magnetic specific heat with triangular lattice model that reveals a similar exchange interaction {as discussed in the next section.}} The low value of the interaction energy scale is attributed to the strongly localized nature of 4\textit{f} orbitals, which is typical in many 4\textit{f}- based frustrated magnets \cite{arh_ising_2022,khatua_spin_2022,guo_magnetism_2019,sala_structure_2023}.

Figure \ref{fig:3}(a) represents the magnetization isotherms obtained at 0.4 K $\leq T \leq$ 30 K.  BYRWO shows paramagnetic behavior for $T \geq$ 4 K, where the spin correlation is negligible. In order to extract \textit{g}, the magnetization isotherms were fitted with the Brillouin function $M/M_s=B_{J}(\textnormal{y})=\left[{\frac{(2J+1)}{2J}\textnormal{coth}\left[\frac{\textnormal{y}(2J+1)}{2J}\right]-\frac{1}{2J}\textnormal{coth}\left[\frac{\textnormal{y}}{2J}\right]}\right]$, in the paramagnetic temperature region $T\geq$ 4 K, where y $=g\mu_\text{B}J(\mu_0H)/(k_\text{B}T)$. The calculated average value of \textit{g} ($\simeq$ 3.12) by assuming $J_\textnormal{eff}=1/2$ is consistent with \textit{g} obtained from the CW fit of the inverse magnetic susceptibility data. {The linear fit to the high field magnetization for the isotherm at 0.4 K gives rise to slope 0.020 $\mu_\text{B}/$T, resulting in the Van Vleck contribution of 0.0113 cm$^3/$mole.} For the magnetization isotherms with $T \leq $ 1 K, at around 3 T, a metamagnetic-like transition appears (see Fig. \ref{fig:3}(b)). The behavior becomes prominent upon decreasing the temperature as demonstrated in Fig \ref{fig:3}(b). It may be noted that at 0.4 K, an anomaly reminiscent of magnetization plateau-like behavior starts at 1 T with magnetization 1.35 $\mu_B$, which is  88 \% of 1.55 $\mu_B$, with which the magnetization tends to saturate. This suggests that around 12 \% of spins are participating in the spin reorientation.
The absence of hysteresis in the magnetic isotherm at 0.4 K during field sweeping, rules out the possibility of a first-order metamagnetic transition.  This can further be supported by $dM/d(\mu_0H)$, which shows a broad hump rather than a sharp peak around the critical field, and becomes more intense as the temperature decreases. While the exact origin of the metamagnetic transition is currently unknown, a similar anomaly in the low-temperature magnetization isotherm has been reported in the {disordered triangular lattice NaYb$_{0.02}$Lu$_{0.98}$Se$_2$ \cite{zhang2025method}. The Re$^{7+}$/W$^{6+}$ site-disorder randomizes the charge environment of Yb$^{3+}$, leading to the random alignments of Yb$^{3+}$ moments. The random alignments of Yb$^{3+}$ moments may lead to cluster of spin singlets. Upon the application of magnetic field, there is a crossover of the spins between the singlet and triplet state \cite{zhang2024observation}, leading to a change in the slope in the magnetization isotherms as observed in  NaYb$_{0.02}$Lu$_{0.98}$Se$_2$ \cite{zhang2025method}}.

\begin{figure*}
    \centering
    \includegraphics[width=1\linewidth]{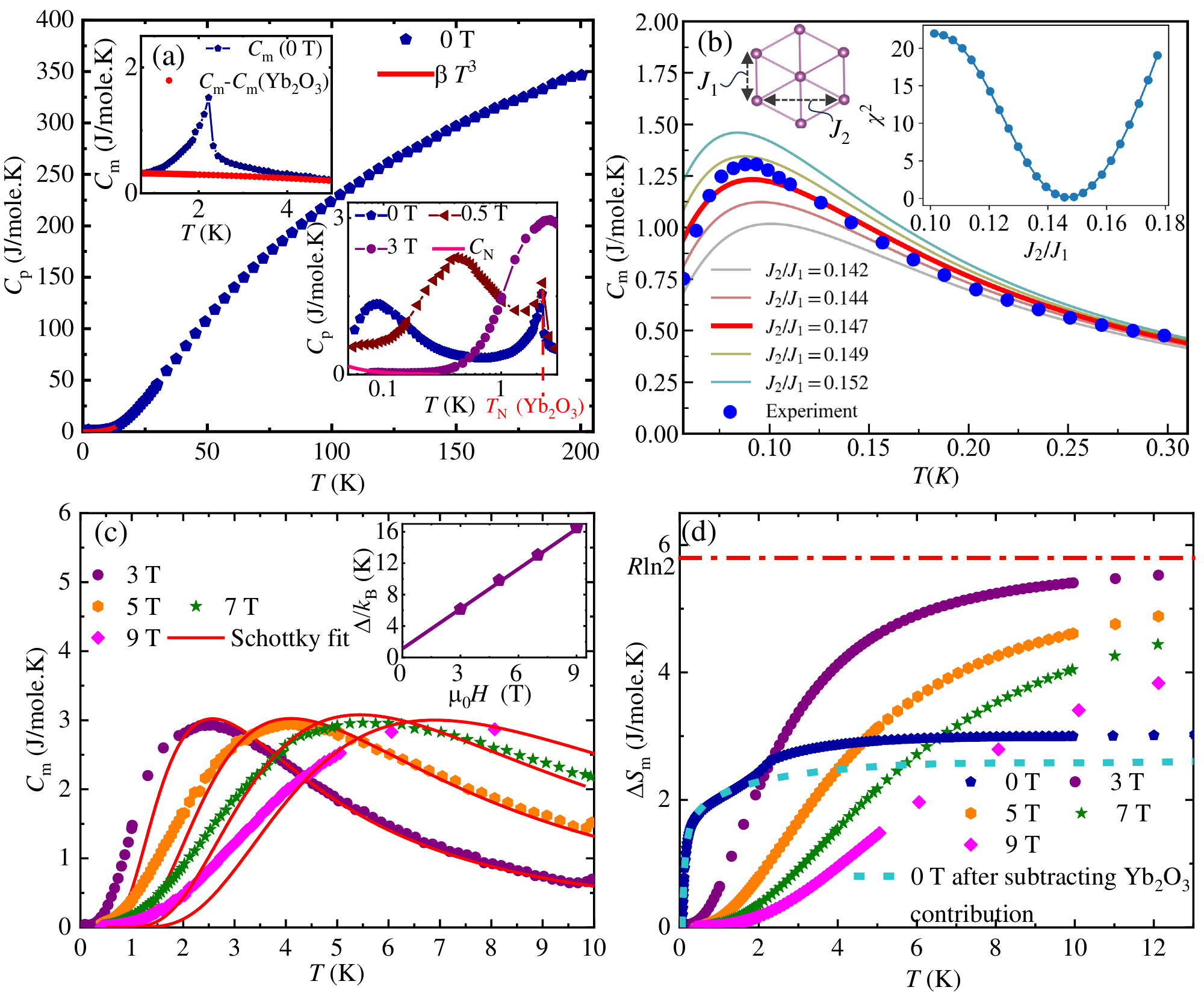}
    \caption{(a) The specific heat at 0 T measured in the range 0.056 K to 200 K. The red solid line represents the $\beta T^3$ fit up to 14 K in order to estimate the lattice specific heat at low temperatures. Bottom inset: the zero field specific heat reveals a transition at 2.2 K, that is due to residual Yb$_2$O$_3$. {The pink line represents the nuclear Schottky ($C_\text{N}$) contribution.}  In the top inset, the curve with red solid spheres denotes a simulated curve in the absence of Yb$_2$O$_3$, by matching the specific heat values below and above the transition. It represents an estimated $C_{m}$ after subtracting the contribution of Yb$_2$O$_3$.~ (b) { Fit of the magnetic specific heat data to the triangular lattice $J_1-J_2$ model, where $J_1$ and $J_2$ represent the nearest and next-nearest neighbor intraplanar interactions, respectively, as depicted in the schematic triangular lattice in the inset. The blue spheres denote the experimental data and the lines correspond to various values of $J_2/J_1$ with $J_1=-0.197$ K. The inset shows the goodness of fit $\chi^2$ for various values of $J_2/J_1$. }(c) The lattice contribution subtracted specific heat at low temperatures. The red solid lines represent the two-level Schottky fit. Inset: The evolution of the Zeeman splitting with magnetic field. The obtained \textit{g} from the slope of the linear fit (purple solid line) is in agreement with the $J_\textnormal{eff}$ = 1/2 nature of the Yb$^{3+}$ moments. (d) Temperature evolution of magnetic entropy in applied magnetic fields. {The cyan dashed line {that saturates at $\approx 2.65$ J/mole.K}, represents the estimated entropy contribution by subtracting the contribution of Yb$_2$O$_3$}. The dotted red line corresponds to entropy \textit{R}ln2, which is expected for $J_\textnormal{eff}$ = 1/2.}
    \label{fig:4}
\end{figure*}

\vspace{-\baselineskip}
\subsection{Specific heat}
The specific heat is an advantageous technique for discerning the density of states in quantum magnets. This, in turn, yields insights into the distinct forms of low-energy excitations, phase transitions and the nature of {spin correlations in} the ground state in frustrated magnets. To unveil the ground state of BYRWO, we conducted the specific heat measurement in zero field in the range 0.056 K$\leq T \leq$ 200 K as shown in Fig. \ref{fig:4}(a). The inset of Fig. \ref{fig:4}(a) illustrates the appearance of an anomaly at 2.2 K in the zero field, which is not associated with the magnetic ordering of BYRWO but attributed to the antiferromagnetic transition arising from a tiny fraction of residual Yb$_2$O$_3$ in the sample {that is unavoidable in some polycrstalline samples }\cite{khatua_magnetic_2024, grusler_role_2023,tokiwa_frustrated_2021}. {However, the main phase of BYRWO does not show a phase transition down to 56 mK, as observed by the zero-field specific heat experiments.} In order to estimate the lattice contribution the specific heat (\textit{C}$_l$) at low temperatures, {the total specific heat} was fitted with the simplest approximation $\beta T^3$ due to the unavailability of non-magnetic analogs and the non-producibility of lattice-specific heat with the Debye-Einstein model. The Debye temperature, related to $\beta$ 
by the expression $\theta_D=\sqrt[3]{12\pi^4Nnk_B/5\beta}$ is estimated to be $\sim 260.01$ K, where $n$, $N$, and $k_\text{B}$ represent the number of atoms in the molecule, Avogadro's number and Boltzmann constant, respectively. {The total specific heat of BYRWO, at 0 T, 0.5 T and 3 T indicate that there is negligible influence of nuclear Schottky effect at 0 T, down to the lowest measured temperature owing to absence of significant remanent heat capacity with increasing the magnetic field at low temperature as observed in NaYbO$_2$ \citep{bordelon2019field}. To get a rough estimate of nuclear Schottky, we fitted the low temperature specific heat at 3 T, due to its lower value of $C_\text{p}$, with $\sim1/T^2$ \cite{sana_magnetic_2023}. The magnetic contribution to the specific heat is estimated by subtracting the lattice and nuclear Schottky contribution from the total specific heat in zero field as presented in Fig. \ref{fig:4}(b)}. {The magnetic specific heat exhibits a broad maximum around 90 mK that indicates short-range spin correlations and suggests a disordered ground state. {The low amplitude of the specific heat broad maximum may be associated with the low interaction strength, which have been observed in other rare earth frustrated magnets \cite{gao_experimental_2019,ranjith_field-induced_2019,baenitz_naybs_2018}. The exchange interaction strength was obtained by fitting the triangular lattice $J_1-J_2$ model to the specific heat with $J_1$ and $J_2$ representing the nearest neighbor and next nearest neighbor exchange interactions, respectively \citep{PhysRevB.109.014425,gonzalez2022ground,pierre2024high}. As the high temperature series expansion (HTSE) interpolation captures only the magnetic contribution to the specific heat \cite{PhysRevB.95.060412,PhysRevB.109.014425}, in the present magnet, the uncertainties in the subtraction of lattice and Yb$_2$O$_3$ may lead to a poor fit at high temperature. Therefore, we considered estimating the exchange strength by fitting the low-temperature $C_\text{m}$ with an exponential fit. For the low-temperature exponential behavior, $C_\text{m}\propto\frac{1}{T^2}\text{exp}(-T_0/T)$, $T_0$ is a function of HTSE coefficients, $J_1$ and $J_2$ as described in \cite{PhysRevB.110.054428,gonzalez2022ground,PhysRevB.101.140403}. For the calculation of $T_0$ the HTSE coefficients of the partition function till $11^{th}$ order were taken into account \cite{pierre2024high} by varying $J_2/J_1$ ratio as depicted in Fig. \ref{fig:4}(b). The best fit results in $J_1\sim$ $-0.197(4)$ K which is close to that obtained from mean-field approximation and $J_2/J_1\sim$ 0.147 (1), indicating that the system stabilizes a spin liquid state down to the lowest measured temperature \cite{PhysRevB.93.144411}.}}

Upon the application of a magnetic field of $\mu_0H =$ 3 T, the transition due to Yb$_2$O$_3$ impurity, observed at 2.2 K entirely disappears (see Fig. \ref{fig:4}(c)). The appearance of the broad maximum with the application of magnetic field that shifts towards higher temperature with higher fields resembles the behavior due to the Schottky anomaly arising from the field-induced splitting of the lowest Kramers doublet with \textit{J}$_\textnormal{eff}$ = 1/2 typically observed in rare-earth based frustrated magnets. To estimate the Zeeman energy gap, the \textit{C}$_p$-\textit{C}$_l$  was fitted with $C_\textnormal{Sch}$, where $C_\textnormal{Sch}$ represents the two-level Schottky expression,
$C_\textnormal{Sch}=fR(\frac{\Delta}{k_\textnormal{B}T})^2\frac{\textnormal{exp}(\frac{\Delta}{k_\textnormal{B}T})}{[1+\textnormal{exp}(\frac{\Delta}{k_\textnormal{B}T})]^2}$, where $\Delta$ is the Zeeman energy splitting of the Kramers doublet ground state, \textit{f} is the fraction of spins of the rare-earth moments contributing to the Kramers doublet states, \textit{R} is the universal gas constant and \textit{k}$_\textnormal{B}$ is the Boltzmann constant. The Schottky fit deviates from the experimental specific heat below 2 K, possibily due to the effect of the impurity or Re/W disorder. The electrostatic field disorder may distort the ideal YbO$_6$ octahedra,  leading to deviations in the energy levels from the expected Zeeman energy spectrum \cite{li_crystalline_2017}. The Schottky fit results $f \sim 0.85$, signifies that the majority of the Yb$^{3+}$ ($J_\text{eff}=1/2$) spins participate in the Kramers doublet state. The inset depicts the evolution of the energy gap as a function of the magnetic field. {The estimated value of \textit{g}, determined from the slope of the linear fit is in reasonable agreement with the magnetization results as expected in $J_\text{eff}=1/2$ spin state of the Yb$^{3+}$ moments}. The intercept of $\sim$ 1.1 K refers to the zero-field splitting (ZFS), which  originates from magnetic anisotropy \cite{valiev2012magnetooptical} or a small admixture of Kramers doublet states. ZFS has been observed in several Yb$^{3+}$ based systems \cite{sana_magnetic_2023,PhysRevB.77.174401}. Additionally the intercept may arise from the uncertainities in the Schottky fit due to the deviation between the Schottky model and the experimental data.

Figure \ref{fig:4}(d) depicts the change in entropy obtained from integrating (C$_p$- C$_l$)/\textit{T} from the lowest measured temperature to 13 K. The zero-field entropy reaches saturation above 5 K, accounting for 52 \% of the total entropy expected for a $J_\textnormal{eff}=1/2$ spin state. {As the entropy includes the contribution from the residual Yb$_2$O$_3$, to estimate the correct entropy contribution from the BYRWO, a smooth curve was simulated to match the specific heat below and above the transition originating from Yb$_2$O$_3$ (see the top inset of Fig.\ref{fig:4}(a)) \cite{tokiwa_frustrated_2021}. The estimated entropy of BYRWO at 0 T {saturates at $\approx$ 2.65 J/mole.K} as depicted by the cyan dotted line in Fig. \ref{fig:4}(d), which accounts for 46 \% of the entropy release, signifying a total of around 6\% of residual  Yb$_2$O$_3$ in the polycrystalline sample. The reduced {specific heat} or entropy release {in zero field}} suggests that the lowest measured temperature may not fully capture the complete {specific heat} {below the broad maximum} or weak exchange interaction and significant frustration is present in this triangular lattice. Similar reduced entropy has also been observed in  several 4\textit{f}-ion  based frustrated magnets Ba$_4$YbTi$_4$O$_{17}$\cite{khatua_magnetic_2024}, Yb$_3$Sc$_2$Ga$_3$O$_{12}$ \cite{sana_magnetic_2023} and KBaYb(BO$_3$)$_2$\cite{tokiwa_frustrated_2021}. With the application of a field of 3 T, the entropy approaches \textit{R}ln2, where 95\% of the entropy is recovered at 11 K.
\begin{figure*}[t]
    \centering
    \includegraphics[width=1\linewidth]{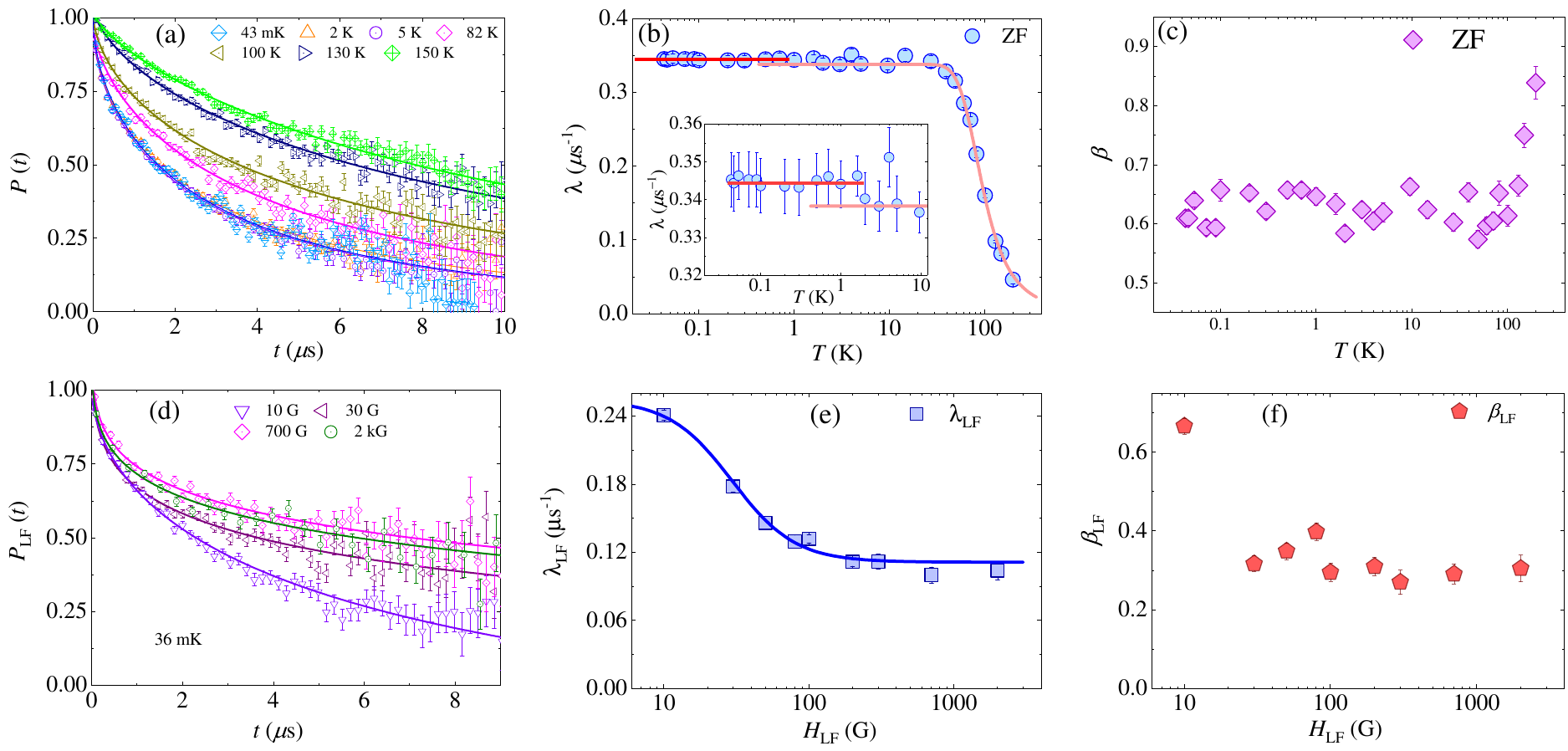}
    \caption{\label{ZF}(a) Time evolution of muon spin polarization at selected representative temperatures in zero field.  (b) Temperature dependence of the muon spin relaxation rate ($\lambda$) where a solid line denotes the thermally activated behavior of  crystal electric field excitations. {The inset depicts the $\mu$SR relaxation rate at low temperature, highlighting the onset of spin correlations below 1 K, ruling out spin freezing down to 43 mK within the experimental time window.} (c) Temperature dependence of the stretched exponent. (d) Time evolution of the muon spin polarization in several longitudinal magnetic fields at 36 mK.  (e) Muon spin relaxation rate ($\lambda_{\rm LF}$)  as a function of LF. The solid blue line represents a phenomenological model given by $\lambda_{\rm LF} (H_{\rm LF}) = \lambda_{\rm LF_{\rm 2D}}(H_{\rm LF}) + \lambda_{\rm 0D}$, which accounts for two-dimensional diffusive spin excitations and zero-dimensional localized spin excitations.   (f)  Stretched exponent ($\beta_{\rm LF}$)  as a function of longitudinal magnetic fields.}
    \label{fig:5}
\end{figure*}

\vspace{-\baselineskip}
\subsection{Muon Spin Relaxation ($\mu$SR)} To shed insights into the ground state and investigate the spin dynamics of strongly localized  $J_{\rm eff} = 1/2$ moments of Yb$^{3+}$ ions arranged on the triangular lattice of Ba$_{4}$YbReWO$_{12}$, ZF $\mu$SR measurements were conducted down to 43 mK, which is much lower than the CW temperature ($\theta_{\rm CW}$=~$-$0.25 ~K).  Figure~\ref{ZF}(a) shows the typical time evolution of muon spin polarization  at selected temperatures in zero-field. The constant background contribution from the silver sample holder in the dilution refrigerator at the M15 station was subtracted by comparing the 2 K data from M15 with that from the M20 station, and the resulting spectra were normalized.  In a long-range ordered magnet, one would typically observe 2/3 oscillating and 1/3 non-oscillating components, resulting from spatial averaging of the  muon spin polarization over all directions in polycrystalline samples. The absence of static magnetic moments  probed at the interstitial muon stopping site within the measured temperature range is confirmed by the following observations: (i) no loss of asymmetry down to 36 mK, (ii) absence of coherent oscillations, and (iii) no crossover of muon spin polarization  at different temperatures in longer timescales. The observed muon spin polarization is best described by a stretched exponential (solid line in Fig.~\ref{ZF}(a))  $P(t) = \exp{[-(\lambda t)^\beta]}$ function, where $\lambda$ is the muon spin relaxation rate and $\beta$ is the stretched exponent that characterizes the distribution of relaxation rates-a typical behavior observed in many frustrated magnets.\\ 
The temperature dependence of $\lambda$ is shown in Fig.~\ref{ZF}(b). It is observed that as the temperature decreases below $150~\text{K}$, $\lambda$ begins to increase and continues this trend down to $30~\text{K}$, below which $\lambda$ remains approximately constant. This increase in $\lambda$ within the temperature range $30~\text{K} \leq T \leq 150~\text{K}$, which occurs well above $\theta_{CW}$, reflects the thermally activated behavior of CEF excitations, a phenomenon commonly observed in rare-earth-based magnets above the Kramers doublet ground state \cite{khatua_magnetic_2024}. Such behavior can be effectively described by the Orbach process, indicating that the relaxation mechanism is likely due to spin-phonon interactions, with an energy barrier governing the relaxation rate that could be linked to the crystal field levels of Yb ions in this triangular lattice. In the absence of any electronic spin correlations at high temperatures, the increase in the muon spin relaxation rate in the temperature range $30~\text{K} \leq T \leq 150~\text{K}$ is attributed to fluctuations of CEF, suggesting a crossover regime from the $J = \frac{7}{2}$ state to the $J_{\rm eff} = \frac{1}{2}$ Kramers doublet state. This crossover can be well described by the phenomenological model relevant for the Orbach relaxation mechanism \cite{arh_ising_2022,reotier_hfinqi_2005} 
\balance
\begin{equation}\label{or}
\frac{1}{\lambda} = \frac{1}{\lambda_0} + \frac{\eta}{\exp\left(\frac{\Delta_{\mu \rm{SR}}}{T}\right) - 1},  \end{equation} 
where $\lambda_0$ accounts for the electron spin fluctuations in the Kramers doublet ground state at $T\rightarrow$ 0, $\eta$ represents the amplitude parameter of relaxation governed by the Orbach process, and $\Delta_{\mu \rm{SR}}$ is the energy gap between the first excited state and the ground state Kramers doublet. The solid line in Fig.~\ref{ZF}(b) is the fit to Eq.~\ref{or}, which yields  {$\lambda_{0}$ = 0.338(2)} $\mu$s$^{-1}$, $\eta$ = 50(3) and $\Delta_{\mu \rm{SR}}$ = 278(4) K. The large value of $\Delta_{\mu \rm{SR}}$ = 278(K) confirms that only the Kramers ground-state doublet is occupied at low temperatures and remains well-separated from the excited states, confirming the $J_{\text{eff}} = \frac{1}{2}$ state of Yb$^{3+}$ ions at low-temperatures. It is worth noting that below 30 K, the temperature-independent value of $\lambda$ is characteristic of electronic spin fluctuations in the CEF ground-state Kramers doublet with weak magnetic exchange interaction between $J_{\rm eff} = 1/2$ moments. Conversely, an additional upturn in $\lambda$ would be observed due to the slowing down of spin dynamics, as observed in YbMgGaO$_{4}$ \cite{pratt_spin_2022}. The observed  stretched exponential behavior of muon spin polarization and the temperature evolution $\lambda$ in zero field suggests the presence of significant spin fluctuations and dynamic spin correlations between Yb$^{3+}$ moments.\\
The estimated parameter $\beta$, shown in Fig.~\ref{ZF}(c),
 remains approximately constant at around 0.63 in the temperature range  0.043 K $\leq$ $T$ $\leq$ 100 K, suggesting a broad distribution of relaxation rates and the presence of multiple muon sites. Above 100 K, $\beta$ shows a trend of increasing towards 1, implying a narrowing of the distribution of relaxation times, likely with relaxation channels dominated by CEF. At low $T$, a distribution of relaxation channels is discernible due to the suppression of CEF contributions. However, this cannot be fully explained by inhomogeneous fluctuations alone, as the precise muon interstitial site is unknown. \\

In order to isolate the dynamic contribution to the relaxation of the muon-spin polarization, $\mu$SR measurements were conducted in several longitudinal fields (LF) at 36 mK. Under LF conditions, the muon spin polarization is primarily governed by the fluctuations of the electronic spins of the $J_{\rm eff} = 1/2$ moments, which are coupled to the implanted muons. Figure~\ref{ZF}(d) shows the time dependence of muon spin polarization in various LF. If the applied LF were to fully suppress the internal correlations, the muon spin polarization would be expected to saturate. However, this is not the case here. Instead, the residual muon spin relaxation observed at 2 kG indicates the presence of dynamic magnetism.
  The solid lines in Fig.~\ref{ZF}(d) represent the fitting curves based on a stretch exponential function  $P_{\rm LF}(t) = \exp\left[-(\lambda_{\rm LF} t)^{\beta_{\rm LF}}\right]$, where the parameters $\lambda_{\rm LF}$ and $\beta_{\rm LF}$ retain their usual meanings. \\ The field dependence of $\lambda_{LF}$
is displayed in Fig.~\ref{ZF}(e). 
At $T = 0.036 ~{K}<<\theta_{\rm CW}$, the field evolution of  $\lambda_{\rm LF}$ directly probes the Fourier transform of the dynamic spin-spin autocorrelation functions, given by  $q(t) = \langle \mathbf{S}_i(t) \cdot \mathbf{S}_i(0) \rangle $, which display either exponential or power-law behavior, depending on the nature of spin dynamics \cite{keren_probing_1996,dunsiger_magnetic_2006,lowe_nuclear_1968}. Furthermore, in the presence of LF, it is well known that  for  a single, well-defined electron-spin fluctuation frequency $\nu_{e} >> \gamma_{\mu}B_{\rm loc}$ an exponential autocorrelation function leads to the Redfield relation 
\begin{equation}\label{red}
\lambda_{\rm LF_{\rm 2D}} (H_{\rm LF}) = \frac{2\gamma_{\mu}^2 B_{\text{loc}}^2 \nu_e}{\gamma_{\mu}^2 H_{\rm LF}^2 + \nu_e^2}
\end{equation}
where $B_{\rm loc}$ represents the width of the fluctuating local fields at the muon sites, $\gamma_{\mu}$ = 135.5$\times$2$\pi$ MHz/T  is the muon gyromagnetic ratio, and $H_{\rm LF}$ is the applied LF.

The field dependence of $\lambda_{\rm LF}$, accurately described by the Redfield relation, has been observed in  various two dimensional frustrated and strongly correlated \begin{table*}[hbt]

    \centering
    \caption{Comparison between the Yb-based triangular lattices and their ground state properties}
\begin{tabular}{ccccc}

    \hline
    \hline
        ~~~~~~~~~~~Material~~~~~~~ & ~~~~~space group~~~~~~ & ~~~~~$\textit{d}_\textnormal{inter}/d_\textnormal{intra}$ ~~~~~~~~ & Low-$\textit{T}$ $\theta_\textnormal{CW}$ \textnormal{(K)}~~ & ~~~~~~~~$\textit{T}_\textnormal{N}$ \textnormal{(K)}~~~~~\\
    \hline
       YbMgGaO$_4$ \cite{li_rare-earth_2015} & $R\Bar{3}m$ & 2.47 &-4   & \textunderscore\\
       NaYbO$_2$ \cite{bordelon_field-tunable_2019} & $R3m$ & 1.64 & -10.3   & \textunderscore\\
       K$_3$\textnormal{Yb(VO}$_4$)$_2$ \cite{voma_electronic_2021} & $P\bar{3}m1$ & 1.29 & -1   & \textunderscore\\
       Rb$_3$\textnormal{Yb(PO}$_4$)$_2$ \cite{guo_crystal_2020} & $P\bar{3}m1$ & 1.43 & -0.056    & \textunderscore\\
       NaBaYb(BO$_3$)$_2$  \cite{guo_magnetism_2019}& $R\Bar{3}m$ & 1.09 & -0.07 & 0.4 \\
       YbBO$_3$ \cite{sala_structure_2023} & $C2/c$ & 1.16 & -0.8  & 0.4 \\
       Ba$_6$\textnormal{Yb}$_2$\textnormal{Ti}$_4$\textnormal{O}$_{17}$ \cite{khatua_magnetic_2024} & $P6_3/mmc$ &~1.26 &-0.49   & 0.077 \\
       Ba$_4$\textnormal{YbReWO}$_{12}$ (\textnormal{\textbf{this work}}) & $R\bar{3}m$ & 1.61 &$-0.25\pm0.01$  & \textunderscore\\
       \hline
       \hline
       
   \end{tabular}
    
    \label{tab:2}
\end{table*}materials and provides insight into the dynamic fluctuations of electronic moments  \cite{arh_ising_2022}. However, our attempts to model the estimated $\lambda_{\rm LF}$ using Eq.\ref{red} were unsuccessful. Instead, the observed behavior is better described by $\lambda_{\rm LF} (H_{\rm LF}) = \lambda_{\rm LF_{\rm 2D}}(H_{\rm LF}) + \lambda_{\rm 0D}$, where the field-independent $\lambda_{\rm 0D}$ refers to zero-dimensional localized spin excitations. The corresponding fit, shown in Fig.~\ref{ZF}(e), yields $B_{\rm loc}$ = 4.98(5) Oe, $\nu_{e}$ = 2.51(2) MHz, and $\lambda_{\rm 0D}$ = 0.111(3) MHz. 
It is worth to note that the obtained $\gamma_{\mu}B_{\rm loc}$ is much smaller than $\nu_{e}$, justifying the use of the Redfield relation for the fast-fluctuating limit.
On the other hand, the finite value of $\lambda_{\rm 0D}$ indicates the existence of zero-dimensional spin fluctuations in the present triangular lattice antiferromagnet \cite{pratt_spin_2022}. A similar scenario has been proposed for two-dimensional triangular lattice antiferromagnets, where the field-dependent relaxation captures spin correlations in the $ab$-plane, while the field-independent term reflects local spin fluctuations uncorrelated with neighboring spins \cite{pratt_spin_2022}. This behavior has also been recently observed in the triangular lattice antiferromagnet $\alpha$-RuI$_3$ \cite{wu_low-temperature_2024}. In the present triangular lattice antiferomagnet, the finite $\lambda_{\rm 0D}$ is attributed to the dipolar coupling between the muon spin and electron spin, which fluctuates rapidly with a rate $\nu_{\rm 0D} >> \gamma_{\mu}B_{\rm loc}$. 
The temperature dependence of $\beta_{\rm LF}$ exhibits a weak dependence on the LF, remaining close to $\beta = 0.32$ for $H_{\rm LF} \geq 30$ G, as shown in Fig.~\ref{ZF}(f). In this regime, the external field is strong enough to suppress certain types of spin fluctuations, leading to stabilized relaxation behavior. In contrast, the value of $\beta_{\rm LF} = 0.65$ for $B_{\rm LF} \leq 30$ G may indicate that the system is more sensitive to slow fluctuations at lower fields.

\vspace{-\baselineskip}

\section{Discussion}
An intricate interplay between the SOC and CEF in 4\textit{f}-based frustrated magnets featuring Kramers doublets with a \textit{J}$_\textnormal{eff}$ = 1/2 state can give rise to a rich landscape of exotic quantum phenomena. The signature of the lowest Kramers doublet with \textit{J}$_\textnormal{eff}$ = 1/2 in the triangular lattice antiferromagnet BYRWO has been uncovered through the reduced magnetic moment and magnetic entropy \textit{R}ln2 in magnetization and specific heat experiments at low temperatures, respectively. The energy gap ($\Delta_\textnormal{CEF}$) of 278 K between the ground-state Kramers doublet and the first excited state, revealed by $\mu$SR, adds further credence to the claim that the ground-state physics is governed by \textit{J}$_\textnormal{eff}$ = 1/2 in this antiferromagnet. The emergence of a broad maximum at 90 mK suggests the short-range correlation among the spins. The YbO$_6$ octahedra, linked to one another through BaO$_{12}$ polyhedra, in contrast to the direct connections of YbO$_6$ octahedra  in NaYbO$_2$ and YbMgGaO$_4$, lead to a relatively weak superexchange interaction $J$ {$\approx-0.197$ K}. In the sub-Kelvin temperature range, the contribution of dipolar interaction is non-negligible, as it has been found in other Yb-based systems. The dipolar interaction was calculated to be around {0.022 K}, by using the expression {$E_\textnormal{dipole}\approx \mu_0 \mu_\textnormal{eff}^2/{4\pi a^3}$, where $\mu_\text{eff}$ is the effective moment} and \textit{a} denotes the nearest neighbor Yb-Yb distance. {In BYRWO, the strength of dipolar interaction is comparable to the intraplanar superexchange interaction between Yb$^{3+}$ moments, suggesting that the ground state results due to the interplay between these two interactions}. The frustration parameter, quantifying the degree of frustration in this triangular lattice antiferromagnet {$f=|\theta_\textnormal{CW}|/T_\textnormal{N}>5$}  indicates frustration in the spin-lattice of BYRWO.  Furthermore, the { 46 \% }entropy released at zero-field down to 56 mK indicates the presence of {frustration induced} strong spin fluctuations and { short-range spin correlations}, which is further indicated by $\mu$SR experiments \cite{khatua_magnetic_2024,sana_magnetic_2023}.

{The absence of magnetic ordering down to 56 mK and a broad maximum in specific heat at $\approx$ 90 mK suggest a disordered ground  state akin to dipolar liquid \cite{bag_realization_2021} in this frustrated magnet. This scenario is corroborated by the absence of oscillations in muon asymmetry,} a characteristic feature of long-range ordering, nor the so-called 1/3 tail down to 43 mK, typical of spin freezing. 
\vspace{-\baselineskip}
\section{Summary}
To summarize, we have successfully synthesized polycrystalline Ba$_4$YbReWO$_{12}$ and conducted investigations of its magnetic ground state through magnetization, specific heat and $\mu$SR techniques. Structural characterization confirms its crystallization in $R\Bar{3}m$ space group, where Yb$^{3+}$ ions constitute a triangular spin lattice. {The reduced magnetic moment obtained from the CW-fit to the low temperature inverse susceptibility suggests the formation of Kramers doublet state. This is further supported by the entropy release calculated from the specific heat measurements in magnetic field.} Furthermore, our $\mu$SR measurements indicate a substantial energy gap of 278 K separating the lowest Kramers doublet ground state from the first excited state. The Curie-Weiss fit to the magnetic susceptibility at low temperatures reveals {a weak} antiferromagnetic exchange interaction {that is supported  by the fit of the zero field magnetic specific heat with the $J_1-J_2$ model with the nearest neighbor exchange interaction  $J_1=-0.197$ K between the $J_\textnormal{eff}=1/2$ moments, which is }{typical of 4$f$ ion based magnets}. The emergence of metamagnetic-like transition are observed in the magnetic isotherms below 1 K, { possibly due to the transition from singlet to triplet state owing to the effect of disorder.} The zero-field specific heat reveals {no signature of magnetic ordering down to 56 mK. The zero-field specific heat shows a broad maximum at ~ 90 mK, which is typical behavior in low-dimensional frustrated magnets suggesting a disordered ground state with short-range spin correlations. The absence of oscillations or a 1/3 tail in muon asymmetry down to 43 mK adds further credence to a disordered ground state in this frustrated $J_\text{eff}=1/2$ triangular lattice antiferromagnet.}
Moreover, gaining a deeper understanding of the nature of magnetic {correlations} by using high-quality single crystals through other microscopic probes such as neutron diffraction and inelastic neutron scattering remain a subject for future investigation. A delicate interplay between competing degrees of freedom in the rare-earth ion-based frustrated triangular lattice antiferromagnet Ba$_4$\textit{R}ReWO$_{12}$ (\textit{R}=rare-earth) offers a promising ground to realize exotic quantum phenomena and establish  realistic Hamiltonian that may broaden our understanding of complex ground states in frustrated quantum magnets.

%%---------------------------------------------
%                     Acknowledgments
%%---------------------------------------------
\section{Acknowledgments}
P.K. acknowledges the funding by the Anusandhan National
Research Foundation (ANRF), Department of Science
and Technology, India through Research Grants. The work at SKKU was supported by the National Research Foundation
    (NRF) of Korea (Grant no. RS-2023-00209121, 2020R1A5A1016518).

\section{Data availability}
The data that support the findings of the current study are available from the corresponding author upon reasonable request.

%%---------------------------------------------
%                     Reference Page
%%---------------------------------------------

%\bibliographystyle{apsrev4-2}
\bibliography{ybrev2}

\end{document}